# Correlation between Trion and Hole in Fermi Distribution in Process of Trion Photo-excitation in Doped QWs

R. A. Suris

*A.F.Ioffe Physico-Technical Institute, Russian Academy of Sciences, 194021 St. Petersburg, Russia*

## *Abstract*

The problem of correlation between a trion and a hole in the electron Fermi distribution created in the process of trion photo-excitation in doped quantum wells is under consideration. The hole in the Fermi distribution appears in the trion creation process consisting of picking of the Fermi Sea electron up by the exciton created in virtual state due to photon absorption. It is demonstrated that the interaction results in formation of a correlated state of the trion and the hole in the Fermi Sea. The state has excitation energy which is less then trion energy minus Fermi energy that can be obtained as a lower edge of trion excitation band using the simple energy conservation low in the picture of independent trion and electrons. The wave function of the correlated state is real and decreases with increase of distance between the trion and the Fermi Sea hole, r, as $1/r^{3/2}$. The wave function can be normalized to unity and it corresponds to correlated state of the trion and Fermi Sea hole. In contrast to this state, the states with excitation energies in the absorption band between the trion energy and the trion energy minus Fermi energy have complex wave functions that decrease as $1/r^{1/2}$. These states correspond to the trion and the Fermi Sea hole that is running away from the trion.

The correlated state described above is supposed to be responsible for the narrow trion absorption line that was observed experimentally.

# 1. Introduction

The problem of optics of doped quantum wells (QW) attracts attention in last decade. The main reason of that is the possibility to observe the manifestation of many-body effects under controllable concentration of carriers in rather wide region of temperature and magnetic field values. Especially, it is of great interest to observe so called trions that are analogous of well-known negatively charged hydrogen atoms, $H^-$, or molecule $H_2^+$. For bulk semiconductors, $H^-$-like complexes were predicted in 1958 [1]. However, their binding energy in bulk semiconductors is too small to make them observable at reasonable temperatures. The reduction of the dimensionality down to 2D was of crucial importance for their observation due to strong increase of the binding energy for interacting confined particles. The first experimental observation of trions in CdTe/CdZnTe quantum wells contained two-dimensional electron gas was done in 1993 [2].

Nowadays, the trions are well studied in QW structures with a two-dimensional electron gas (2DEG) of low density in QWs based on $A^{II}B^{VI}$ as well as $A^{III}B^{V}$ semiconductor compounds. Negatively charged- and positively charged exciton complexes with heavy-holes and light-holes were found experimentally [3-6]. In addition to singlet trion states, triplet states were observed in high magnetic fields [5].

However, up to now the properties of the trion states in 2DEG are far from complete understanding. In [7] we presented a simple theory of optical spectra near trion and exciton resonances. The purpose of the paper is to develop the theory and to give a picture of the photo-excited trion states in weakly doped QWs. Special attention will be paid to analysis of the nature of the state corresponding to a sharp low-energy peak of the trion absorption. We will show that the peak arises due to strong correlation between the trion and the hole in the electron Fermi Sea. The hole is created due to picking Fermi Sea electron up by exciton in virtual state in process of trion formation.

# 2. QW Susceptibility

## 2.1. Main Processes and Approximations

Our task is to calculate optical spectrum of QWs modified by a dilute 2D electron gas. For this consideration, we will be assumed that the sheet electron concentration $n_e$ is so small that $n_e a_{tr}^2 << 1$ (here $a_{tr}$ is the trion radius). It means that i.e. the Fermi energy of 2DEG, $F$, is much smaller than the trion binding energy $E_{tr}$, $F << E_{tr}$. Under this condition, the electron gas does not modify the wave functions of trion and, of course, exciton states and does not affect their binding energies. Here we will limit ourselves with consideration of zero temperature, $T$ = 0 K.

A trion contribution to optical spectrum is essential in the energy region near the exciton resonance because the trion binding energy is much smaller than the binding energy of the exciton.

A naive picture of trion photo-excitation is as follows. A photon creates an exciton in a virtual state, which picks a background electron up forming a trion in final state (see diagram (1) below). It is clear that the photon energy measured relative to the exciton energy should be $\Omega = -E_{tr} - \varepsilon$ (we use units $\hbar \equiv 1$). Here $\varepsilon$ is energy of the electron and $\Omega$ is the photon energy minus the excitation energy of unperturbed exciton. Therefore, due to the Fermi statistics for electrons one can assume that the photon energy required for trion creation belongs to the energy band ranging from $-E_{tr}$ to $-E_{tr} - F$ and absorption spectrum is:

$$abs \propto \int_0^F d\varepsilon \delta\left(\Omega + \varepsilon + E_{tr}\right)$$

However, we will see further that this picture is not complete and we will find that a sharp absorption peak will appear slightly below this band.

Two types of processes caused by exciton-electron interaction are important for our considerations:

i) Trion formation involving an intermediate state "exciton + electron":

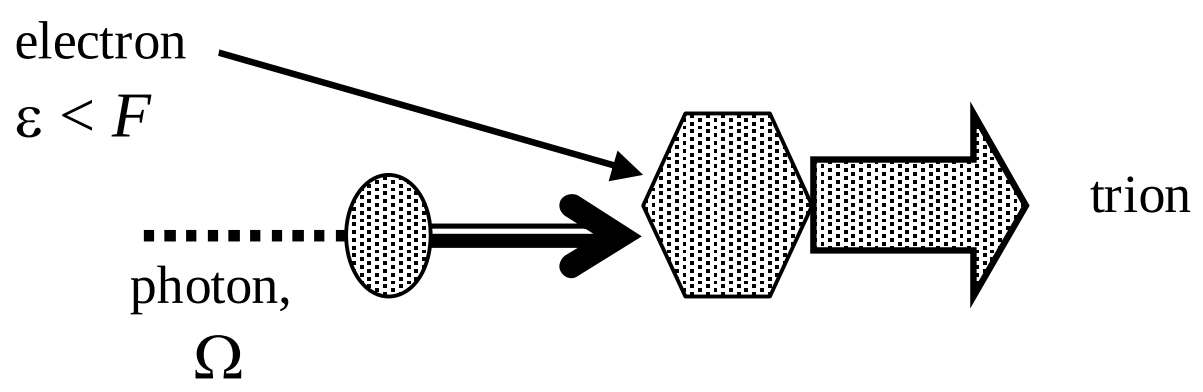

(1)

exciton

ii) Scattering of a photo-excited exciton on a background electron:

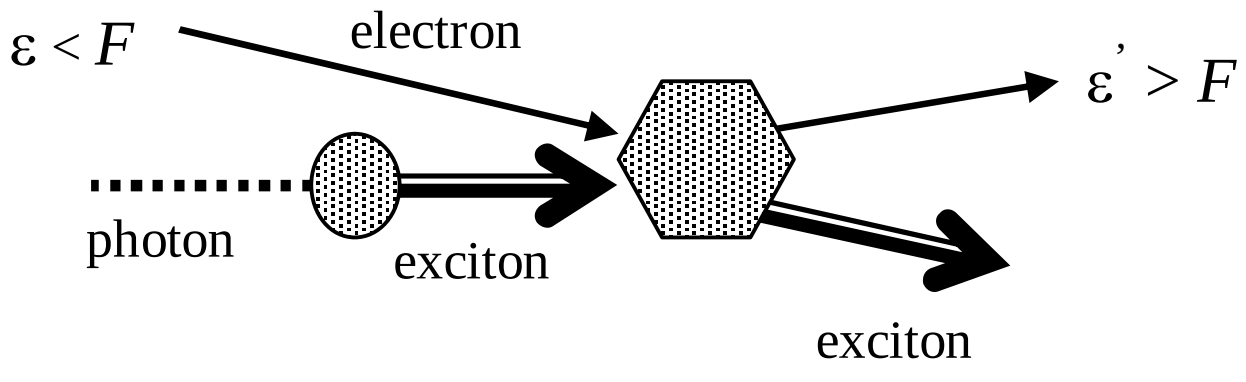

(2)

Because of ε' > ε, processes (2) result in high-energy tail of the exciton resonance.

Since the trion binding energy $E_{tr}$ is much less than the exciton binding energy $E_{exc}$, and as we are interested in the energy range close to the exciton and trion resonances, these processes should be the most effective. The reason is that the processes have an exciton as intermediate state. As a result a resonant denominator from the exciton Green's function appears in their amplitudes. So, these amplitudes are by factor ($E_{exc}$ /$E_{tr}$) larger than the amplitudes of all other processes.

Besides, we should consider of course the process of exciton creation:

(3)

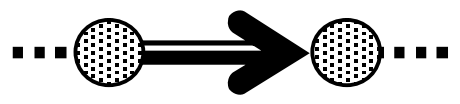

## 2.2. *Diagrams for QW susceptibility*

The complete information on optics of the QW gives its susceptibility.

The susceptibility of the QW in the vicinity of the exciton resonance is given by the sum of the following diagrams:

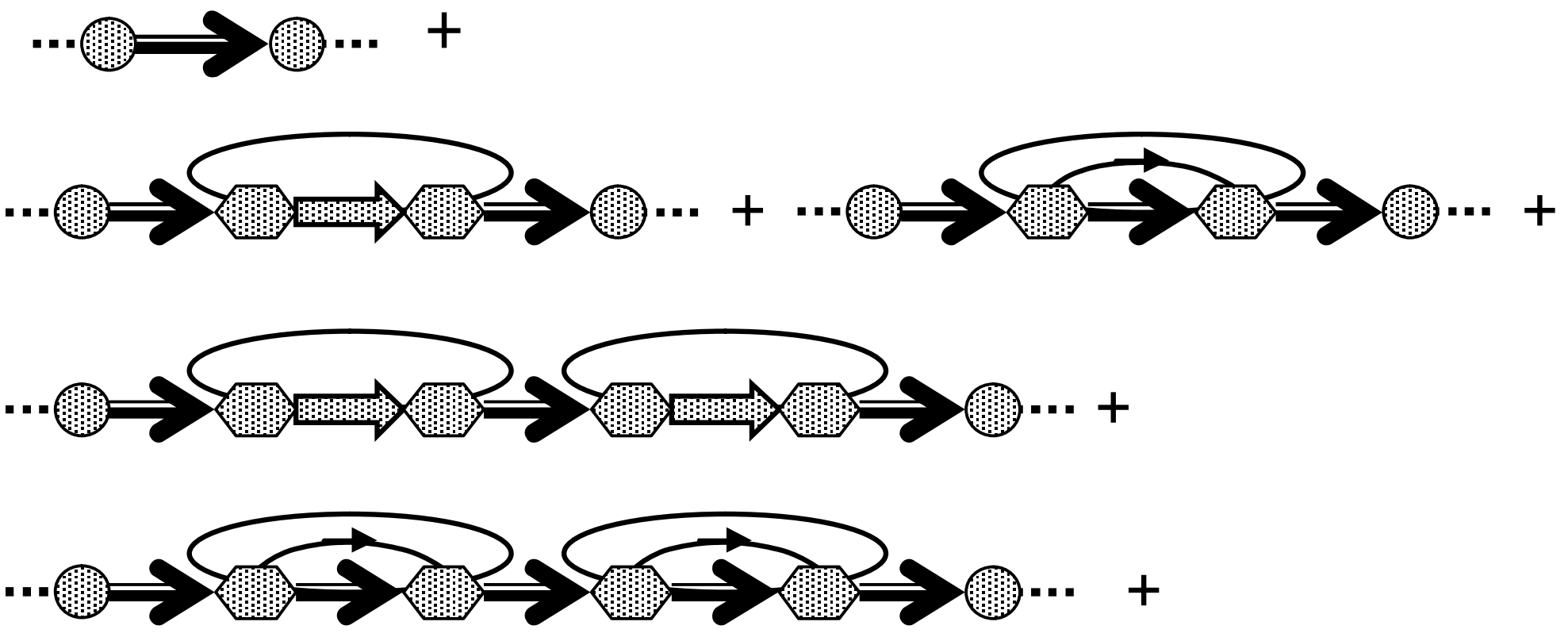

(4)

The first diagram gives simply the exciton susceptibility. Other diagrams account for mutual multiple transformations of exciton into trion and back and for exciton-electron scattering. Despite of the fact that the higher order diagrams are small, we have to sum up all of them because of a resonance contribution from the denominator in the exciton Green function.

The two diagrams in the second line could be represented in a common form as:

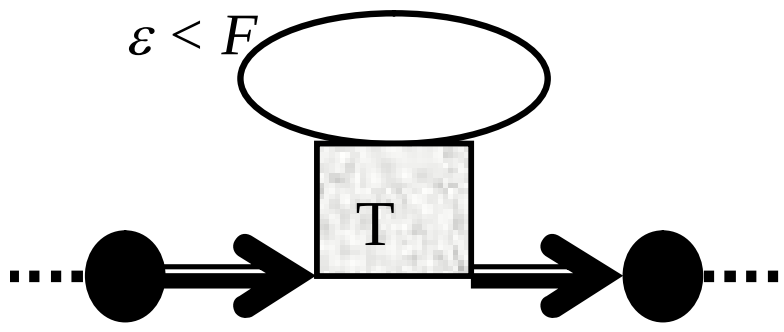

$$(5)$$

The square

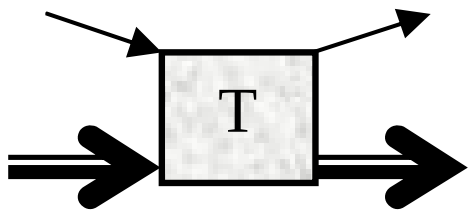

is the amplitude of exciton-electron scattering.

In the terms of the scattering amplitude, the diagrams for the QW susceptibility can be presented in the following form:

$$\chi(\Omega) =$$ 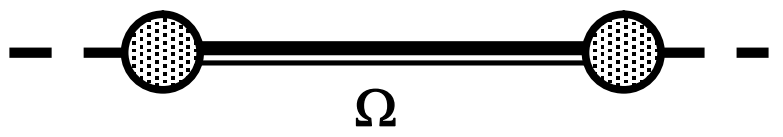

$$+ \quad \left(\propto \frac{1}{\Omega}\right)$$

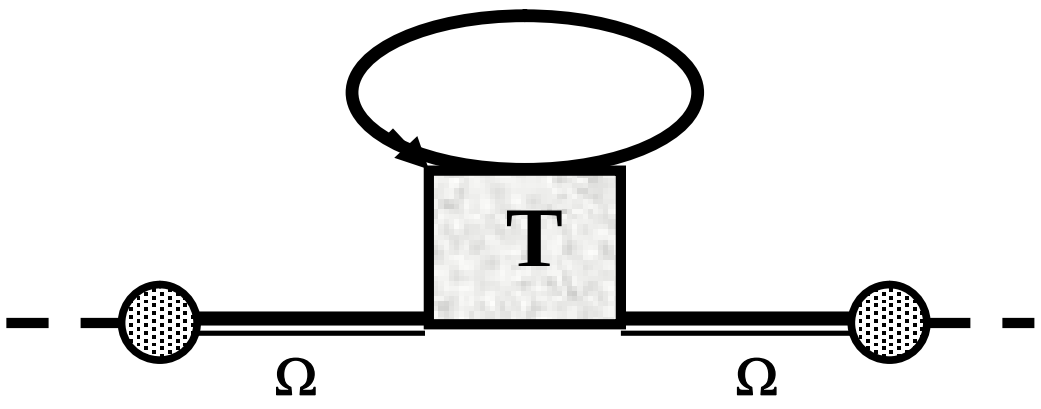

$$+ \quad \propto \left(\frac{1}{\Omega^2}\right)$$

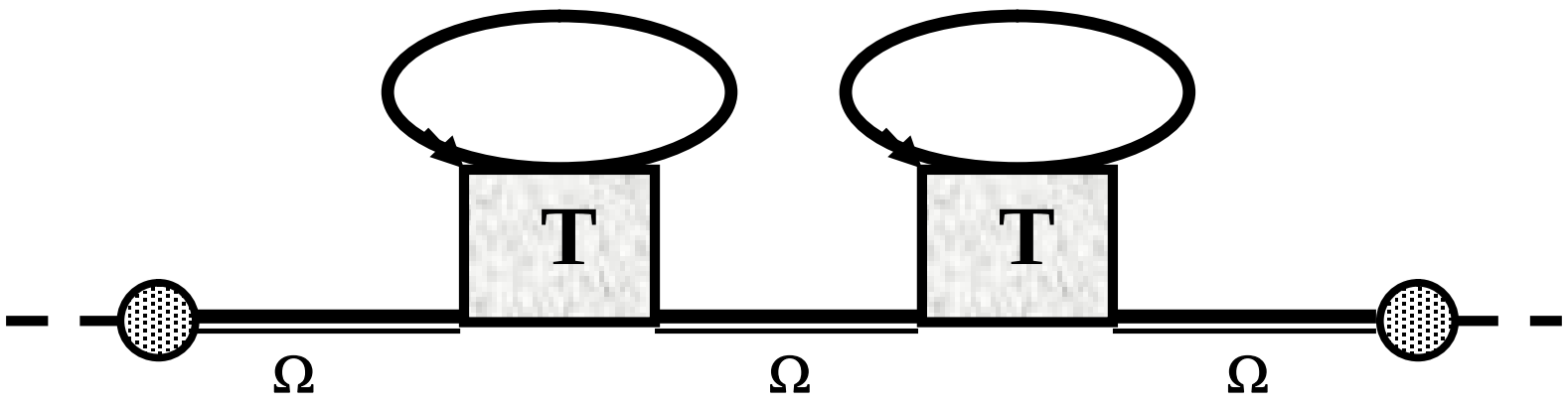

$$+ \quad \left(\propto \frac{1}{\Omega^3}\right)$$

$$\bullet \bullet \bullet =$$

$$= \frac{1}{\Omega + i\gamma - \Pi(\Omega)} \qquad (6)$$

Here we introduce self-energy operator

$$\Pi(\Omega) = \int_0^F \frac{d\varepsilon}{2\pi}$$ 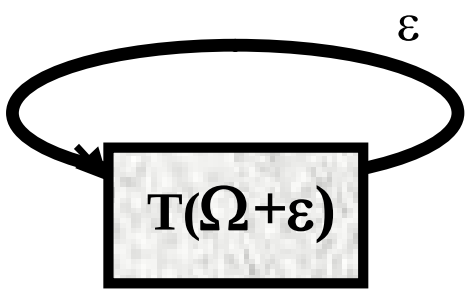 (7)

Let us emphasize again that, despite of the fact that the higher order diagrams are small as corresponding power of concentration, we have to sum up all of them because of a resonance contribution from the exciton Green function that is proportional to 1/Ω.

We supposed the scattering amplitude independent on electron and exciton wave numbers for small relative electron-exciton momentum. That is true as far as Fermi momentum of the electron 2D gas is much smaller then the trion radius as we assumed at very beginning. However, we have to take into account the difference in amplitudes of electron-exciton scattering in singlet and triplet states. Therefore from (7) we obtain:

$$\chi_{\uparrow}(\Omega) = \chi_{\downarrow}(\Omega) = \frac{\left|D_{CV}\right|^2 \Psi(0)^2}{\Omega + i\gamma - \int_0^F \left(\frac{1}{2} T_{sngl}\left(\Omega + \varepsilon\left(1 - \frac{m_e}{2m_e+m_h}\right)\right) + \frac{3}{2} T_{trpl}\left(\Omega + \varepsilon\left(1 - \frac{m_e}{2m_e+m_h}\right)\right)\right)\left(1 + \frac{m_e}{m_e+m_h}\right) d\varepsilon} \quad (8)$$

Here $D_{CV}$ is a conducting band – valence band dipole matrix element and $\Psi(0)$ is electron-hole relative motion wave function of the exciton at zero electron-hole distance.

Arrows in Eq. (8) indicate the electron spin orientation in the exciton defined by the circular polarization of incident photon. $T_{sngl}$ and $T_{trpl}$ are amplitudes of an electron-exciton scattering resulting in a singlet- and triplet state, respectively. Factors $1/2$ and $3/2$ account for the statistical weight of these states. We assume in Eq. (6) that the 2D electrons have no spin polarization. In case of such a polarization (e.g. induced by external magnetic fields) the polarizabilities $\chi_{\uparrow}(\Omega)$ and $\chi_{\downarrow}(\Omega)$ are different, which means that the QW optical properties are different in two circular polarizations of light. The second term in the brackets $(1 - \frac{m_e}{2m_e+m_h})$ and $\left(1 + \frac{m_e}{m_e+m_h}\right)$ account the exciton and trion kinetic energies.

## *2.3. Scattering amplitudes and the trion singularity of the QW susceptibility*

Let us analyze now the scattering amplitudes in Eq. (5). It is known that at zero or weak magnetic fields the trion singlet state is bound whereas its triplet state is unbound. Consequently, the amplitude $T_{sngl}(\Omega)$ has a pole at $\Omega = -E_{tr}$ , whereas the amplitude $T_{trpl}(\Omega)$ has no pole in the energy range $|\Omega| \propto E_{tr}$. Due to the short-range character of the exciton-electron interaction only the amplitude of the *S*-scattering will be calculated at energies less than the trion binding energy. In this case one can show that for $\Omega << E_{Exc}$ (see Appendix)

$$T_{sngl}(\Omega) = \frac{2\pi}{\overline{m}} \bigg/ Ln\left(\frac{E_{tr}}{-\Omega}\right) \qquad \text{and} \qquad T_{trpl}(\Omega) \approx \frac{2\pi}{\overline{m}} \bigg/ Ln\left(\frac{E_{exc}}{-\Omega}\frac{E_{exc}}{E_1}\right) \tag{9}$$

Here, $E_1$ is an energy of the order of $E_{tr}$, and $\overline{m}$ is the reduced effective mass of the system "exciton + electron", $1/\overline{m} = 1/m_e + 1/(m_e + m_h)$. $E_B$ is a 2D exciton Rydberg.

The imaginary parts of the amplitudes $T_{sngl}(\Omega)$ and $T_{trpl}(\Omega)$ are nonzero at $\Omega > 0$ only. This means that the imaginary part of the integral in Eq. (5) is equal to zero if $\Omega < -E_{tr} - F\left(1 - \frac{m_e}{2m_e + m_h}\right)$. One can be convinced that the integral is positive and diverges logarithmically when $\Omega$ approaches to $-E_{tr} - F\left(1 - \frac{m_e}{2m_e + m_h}\right)$ from the low energy side. As a result, the polarizabilities $\chi_\uparrow(\Omega)$, $\chi_\downarrow(\Omega)$ have a pole at a real value of the energy, $\Omega < -E_{tr} - F\left(1 - \frac{m_e}{2m_e + m_h}\right)$. As we will see in the next section, this pole corresponds to a correlated state of the trion and a hole in the Fermi Sea, which appears due to the capture of an electron by an exciton in the process of the trion formation.

The pole is clearly seen as a sharp line in Fig. 1 which gives dependence of log(Im($\chi$)) on the photon energy, $\Omega$, at Fermi energies $F = 0$ and $F = 0.06$. Fig. 2 illustrates evolution of the spectrum with increase of $F$ from 0 up to 0.06. Fermi energies are measured in the units of the 2D exciton binding energy. In these calculations we assumed the following values of parameters: $E_{tr}/E_{exc} = 0.15$, $m_h/m_e = 3$.

One can see that the gap between the pole and the left edge of the trion band, $-E_{tr} - F\left(1 - \frac{m_e}{2m_e + m_h}\right) < \Omega < -E_{tr}$, increases with increase of Fermi energy. One can see also the high-energy tail to the right of the exciton peak. The tail is due to the combined processes presented in diagram (2) and it increases with increase of the electron concentration.

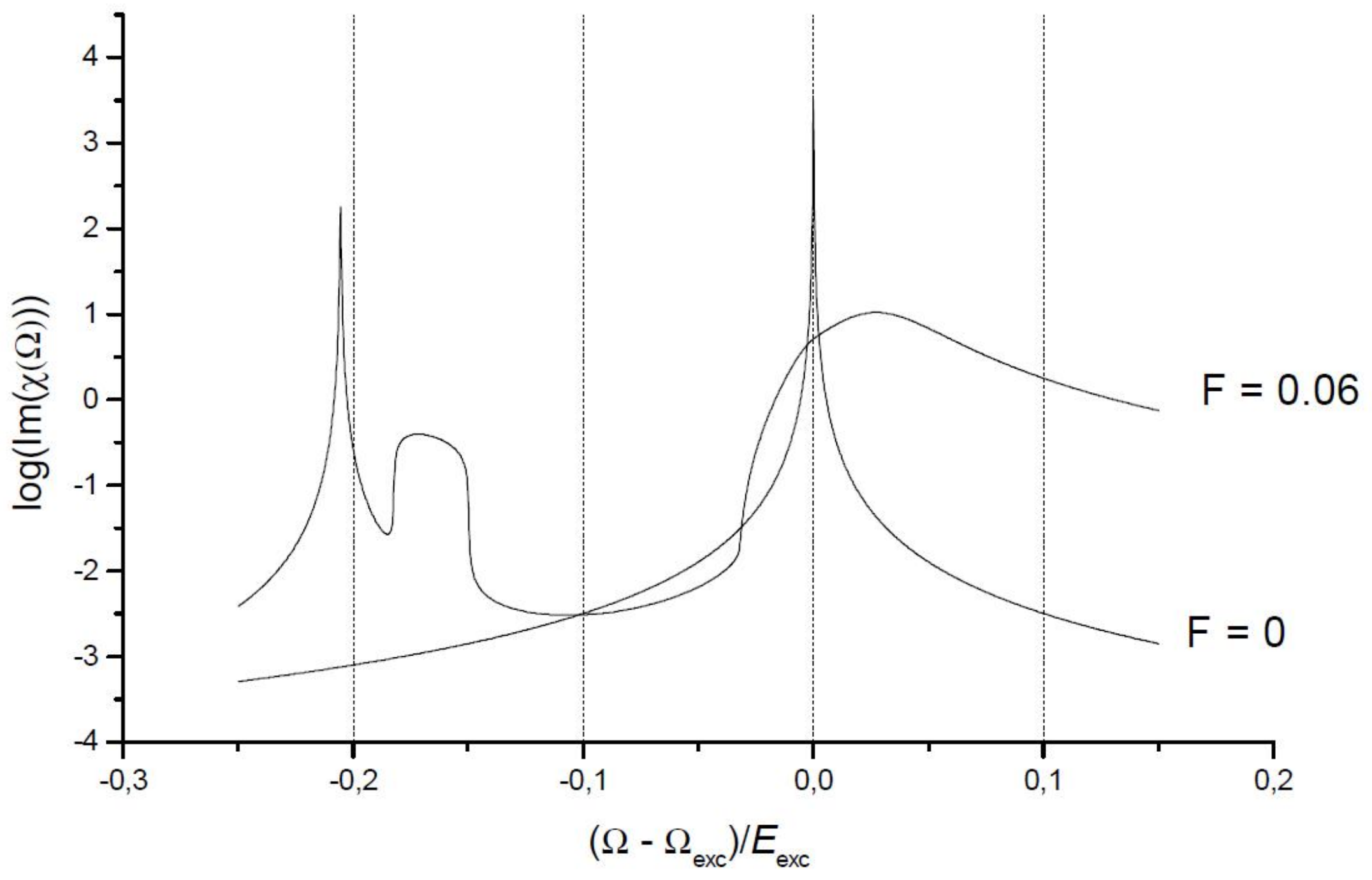

Fig.1. *Dependence of log(Im(χ)) on the photon energy, Ω, at Fermi energies F = 0 and F = 0.06. ( $E_{tr}$ = 0.15, $m_h/m_e$ = 3). All energies are measured in $E_{exc}$ units.*

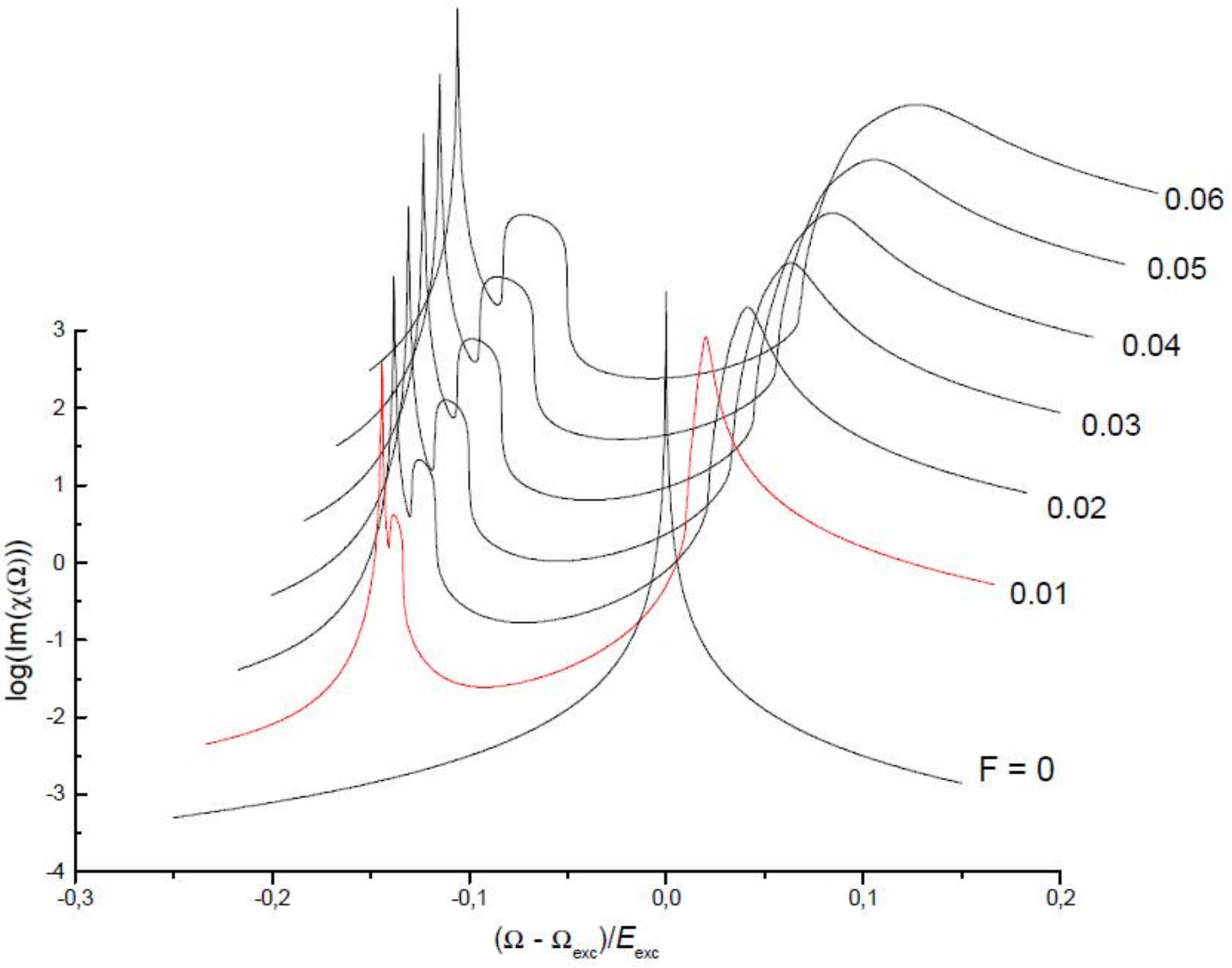

Fig. 2. *Evolution of the spectrum with increase of Fermi energy F from 0 up to 0.06 (in $E_{exc}$ units) ( $E_{tr}$ = 0.15 (in $E_{exc}$ units), $m_h/m_e$ = 3).*

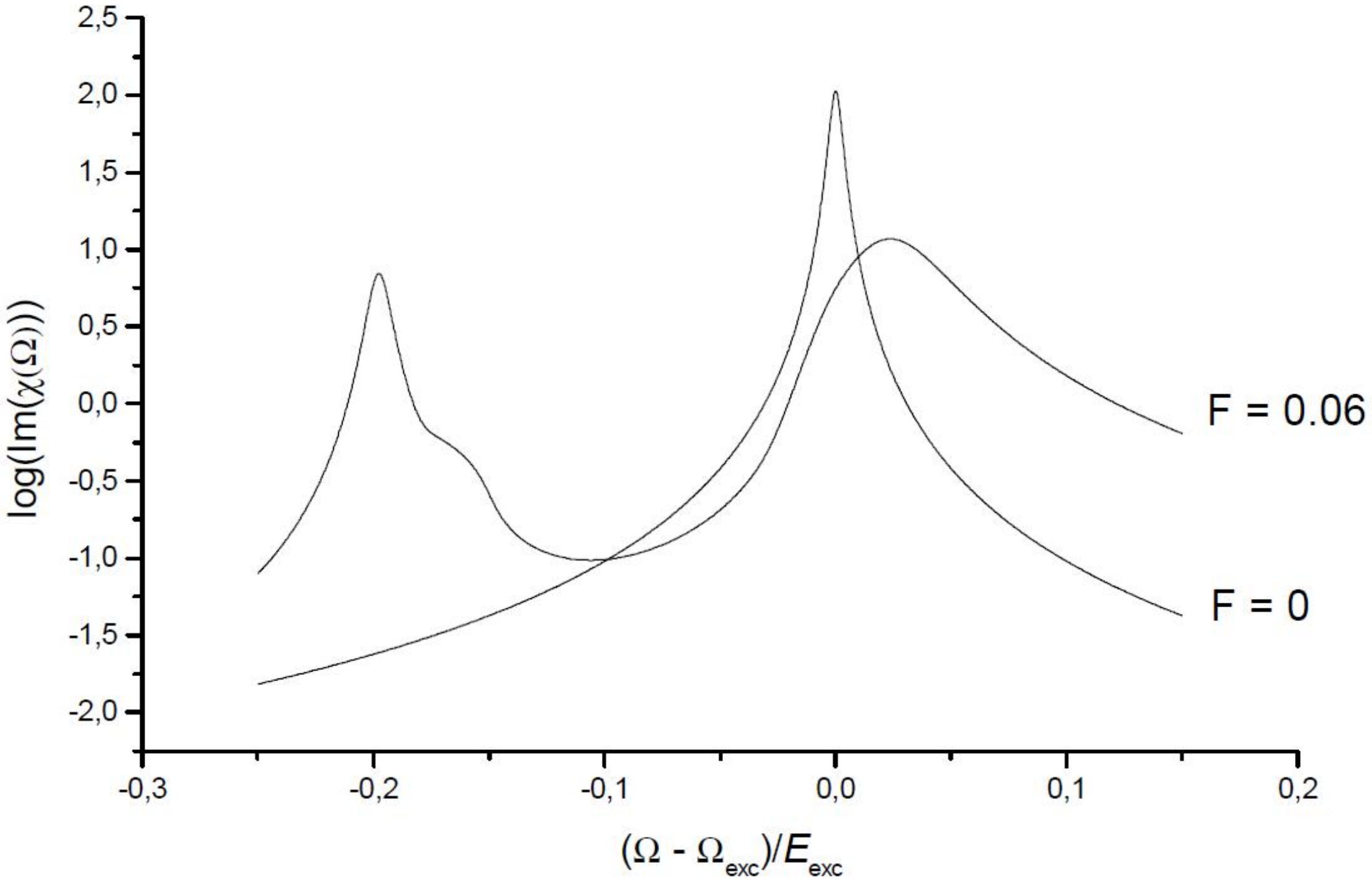

Fig.3. *Evolution of the spectrum with increase of Fermi energy F from 0 up to 0.06 (in $E_{exc}$ units) regarding the exciton and electron dmping: $\gamma_{exs}$ = $\gamma_{el}$ = 0.003 (in $E_{exc}$ units) ( $E_{tr}$ = 0.15 (in $E_{exc}$ units), $m_h/m_e$ = 3).*

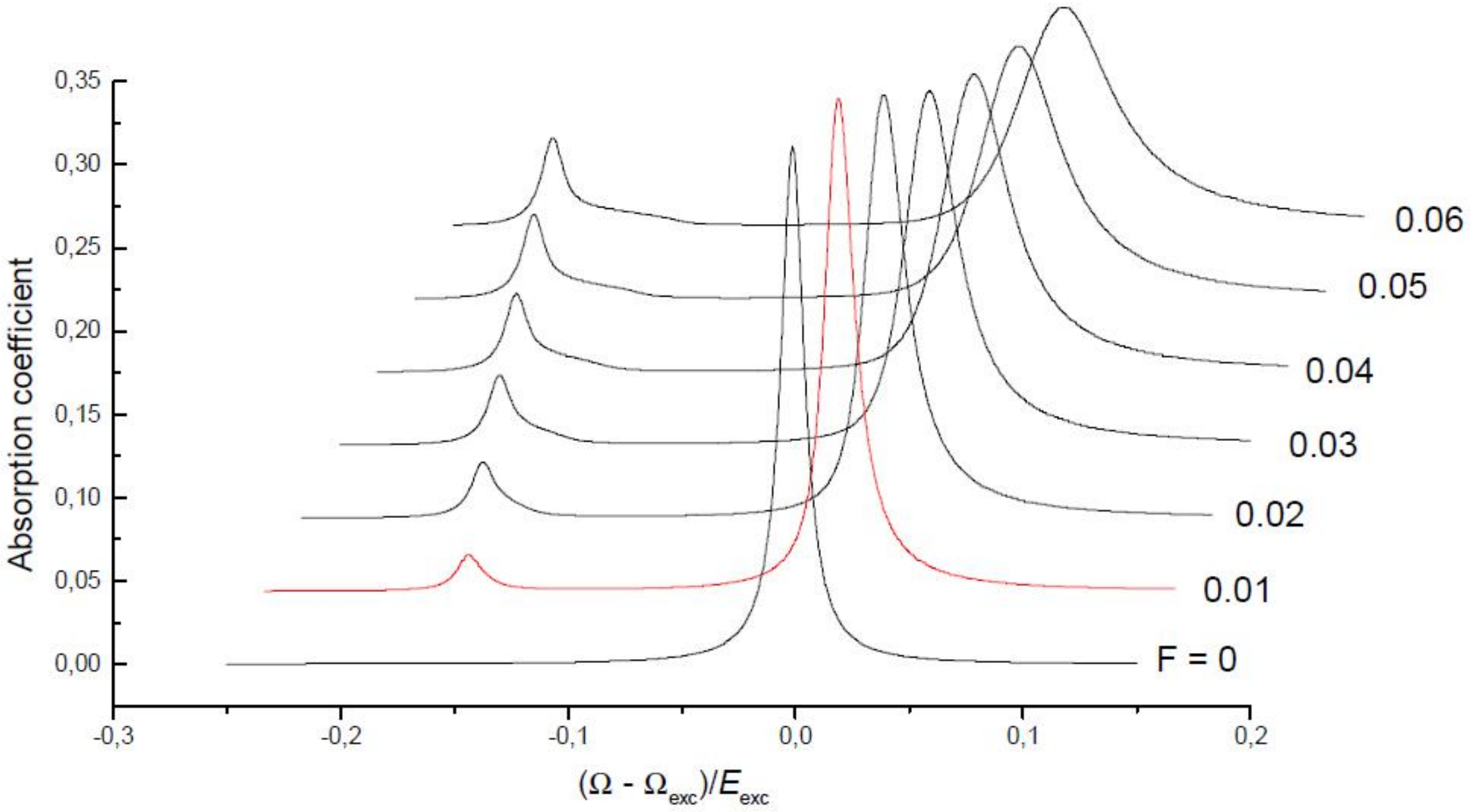

Fig.4. *Dependence of QW absorption coefficient on frequency and Fermi energy for normally incident light ($E_{tr}$ = 0.15, $\gamma_{exs}$ = $\gamma_{el}$ = 0.003 (in $E_{exc}$ units), refraction index = 2.65, Wave vector = 1.4 mkm $^{-1}$, distance of QW from the crystal surface = 0.5 mkm, oscillator strength = 0.005).*

If we take into account the electron and exciton scattering in simplest way, introducing damping parameters for both of them, the peak on the left of the trion band becomes wider and the gap between it and the band disappears (Fig. 3). For this example, we took the damping parameters as $\gamma_{exs} = \gamma_{el} = 0.003$ (in $E_{exc}$ units). The results presented in Fig. 1 and 2 were calculated at $\gamma_{exs} = \gamma_{el} = 10^{-4}$.

For normally incident light, dependence of QW absorption coefficient on frequency and Fermi energy is presented in Fig. 4 at the same values of mass, $E_{tr}$ and damping. We used in this example the following values of parameters:

Refraction index = 2.65,

Wave vector = 1.4 mkm $^{-1}$,

Distance of QW from the crystal surface = 0.5 mkm,

Oscillator strength = 0.005·

The absorption coefficient is obtained be means of solution of the simple electrodynamics problem supposing the QW width much less then the light wavelength.

## 3. *Nature of the singularity*

In this section, we will give the simpl picture of the state responsible for the left-hand delta-function peak in the spectrum described above.

Due to very small value of the trion binding energy, we have to consider a mixture of the trion and exciton wave functions. Mixing of the states is because of interaction with Fermi Sea electrons. The following diagram describes the process:

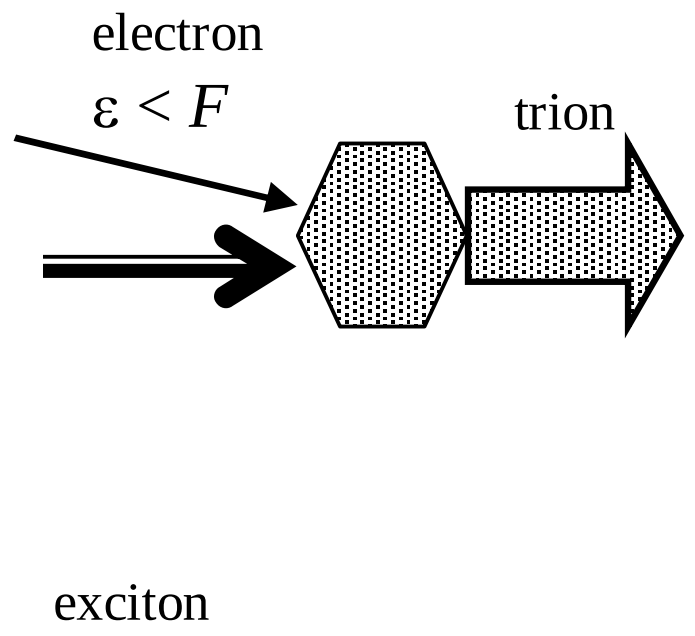

(10)

Actually, we deal with the exciton on the background of Fermi Sea and with the trion on the background of Fermi Sea with a Fermi Sea hole. This can be illustrated as it is shown in Fig. 5. The hole arises due to picking of an electron of Fermi Sea up by the exciton in the virtual state to create the trion (see the diagram above). Therefore, we have to consider the four-particle state:

Trion (two electrons + a valence band hole) + Hole in Fermi Sea.

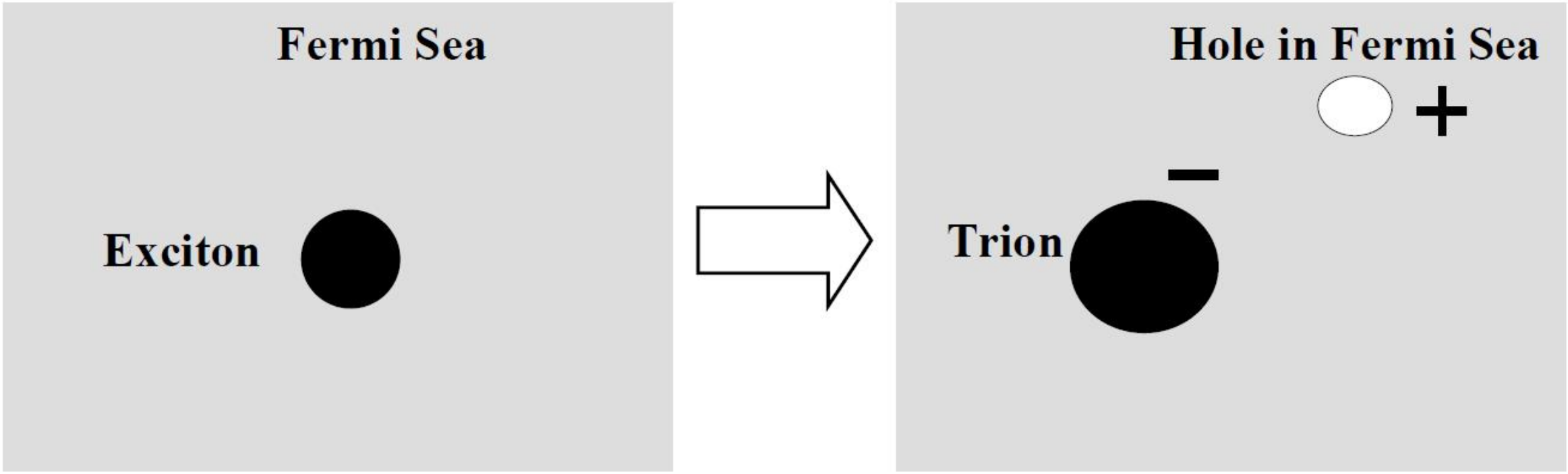

Fig. 5. *Schematic illustration of the trion creation on the Fermi Sea background.*

The wave function (WF) of the mixture of the state with the exciton state is

$$|\Psi\rangle = Exc|\mathbf{Exc}\rangle + \int_{\mathbf{k}} Tr(\mathbf{k})|\mathbf{Tr_k}\rangle \tag{11}$$

The WF of the exciton on the background of Fermi Sea is

$$|\mathbf{Exc}\rangle = \int_{\mathbf{R},\mathbf{r}} \varphi_{exc}(\mathbf{r}-\mathbf{R})\, a_s^+(\mathbf{r})\, b_\sigma^+(\mathbf{R})|\mathbf{\Phi}\rangle . \tag{12}$$

Here $|\mathbf{\Phi}\rangle$ is Fermi Sea WF, $a_s^+(\mathbf{r})$ is the creation operator of the electron with spin S at the point with the coordinate **r**, $b_\sigma^+(\mathbf{R})$ is the creation operator of the hole with spin σ at the point **R**. The function $\varphi_{exc}(\mathbf{r}-\mathbf{R})$ is the WF of the electron-hole relative motion in the free exciton with zero momentum.

WF of the trion + the hole with wave number **k** in Fermi Sea is

$$|\mathbf{Tr_k}\rangle = \int_{\mathbf{R},\mathbf{r}} \varphi_{tr}(\mathbf{r}_1-\mathbf{R},\mathbf{r}_2-\mathbf{R})\, a_{-s}^+(\mathbf{r}_1)\, a_s^+(\mathbf{r}_2)\, b_\sigma^+(\mathbf{R})\, a_{\mathbf{k},-s}|\mathbf{\Phi}\rangle \tag{13}$$

Here $\varphi_{tr}(\mathbf{r}_1-\mathbf{R},\mathbf{r}_2-\mathbf{R})$ is the free trion WF and $a_{\mathbf{k},-s}$ is the annihilation operator of Fermi Sea electron with momentum **k** and spin –S[1]. Due to annihilation operator $a_{\mathbf{k},-S}$ in Eq. (13), the trion WF is zero at $|\mathbf{k}| > k_F$ , where $k_F$ is the Fermi momentum.

[1] Let us remind that operators $a(\mathbf{r})$ and $a_{\mathbf{k}}$ are connected by Fourier-transformation.

Generally, the approximate states (12) and (13) are not orthogonal. However, in the low-density limit, $n_e a_{tr}^2 << 1$, we are eligible to disregard the point.

The equation for amplitudes *Exc* and $Tr(\mathbf{k})$ are

$$\begin{aligned} &\Omega\, Exc - \int_{|\mathbf{k}|<k_F} g(\mathbf{k}) Tr(\mathbf{k}) = 0 \\ &(\Omega + E_{tr} - (-\varepsilon(\mathbf{k}))) Tr(\mathbf{k}) - g(\mathbf{k}) Exc = 0 \end{aligned} \tag{14}$$

Here $-\varepsilon(\mathbf{k}) = -\mathbf{k}^2/2m_{el}$ is the energy of the Fermi Sea hole with momentum –**k**. Function $g(\mathbf{k})$ is the vortex part in the diagram (13). The function can be expressed in terms of the trion WF.

Equation for eigenvalues of the system is

$$\Omega - \int_{|\mathbf{k}|<k_F} \frac{g(\mathbf{k})^2}{\Omega + E_{tr} + \varepsilon(\mathbf{k})} = 0 \tag{15}$$

Actually, this equation is the same as the equation for pole of the $\chi$ (Eq. (8)) obtained by expanding the first term in self-energy operator $\Pi$ near the trion pole in the limit $m_e/m_h \to 0$.

Due to our initial assumption, $k_F << a_{tr}$, we can neglect the **k**-dependence of $g(\mathbf{k})$ and find the value of $g(\mathbf{0})$ from Eqs. (8) and (9). Therefore, we have the dispersion equation in the simple form

$$\Omega - \int_0^F \frac{E_{tr}}{\Omega + E_{tr} + \varepsilon} d\varepsilon = 0 \tag{16}$$

or

$$\frac{\Omega}{E_{tr}} - \ln\left( \frac{1 + \frac{\Omega}{E_{tr}} + \frac{F}{E_{tr}}}{1 + \frac{\Omega}{E_{tr}}} \right) = 0 \tag{16'}$$

The dispersion equation should be considered only at $\Omega < -E_{tr} - F$. For example, the solution of the equation for $F$ = 0.06 and $E_{tr}$ = .15 (in the units of the 2D exciton energy, $E_{exc}$) $\Omega = -E_{tr} - F - 0.017$. This pole is clearly seen as a sharp line in Fig. 1. In the energy region $-E_{tr} - F < \Omega < -E_{tr}$ we deal with continuous spectrum.

Using Eq. (14), the trion amplitude $Tr(\mathbf{k})$ can be presented as

$$Tr(\mathbf{k}) = \frac{g(\mathbf{k})}{\Omega + E_{tr} + \varepsilon(\mathbf{k})} Exc \tag{17}$$

The spatial correlation of the photo-created trion and the hole in Fermi Sea is given by Fourier-transformation of the function $Tr(\mathbf{k})$:

$$\Phi(\mathbf{r}) \propto \int_{|\mathbf{k}|<k_F} Tr(\mathbf{k}) e^{i\mathbf{k}\cdot\mathbf{r}} \frac{d^2\mathbf{k}}{(2\pi)^2} \tag{18}$$

Actually, this function WF describes the relative motion of the Fermi Sea hole and the trion.

The behavior of the function radically different in the two energy regions:

1. $\Omega < -E_{tr} - F$

2. $-E_{tr} - F < \Omega < -E_{tr}$ .

In the first region of energies the function (18) is real and, oscillating, it approaches to zero in the limit $|\mathbf{r}| \to \infty$ as $|\mathbf{r}|^{-3/2}$ where $|\mathbf{r}|$ is a distance between the Fermi Sea hole and the trion. Let us stress that, due to the strong decrease with distance, the function can be normalized to unity. *It corresponds to the Fermi Sea hole localized around the trion*. The behavior described above is illustrated in Fig. 6.

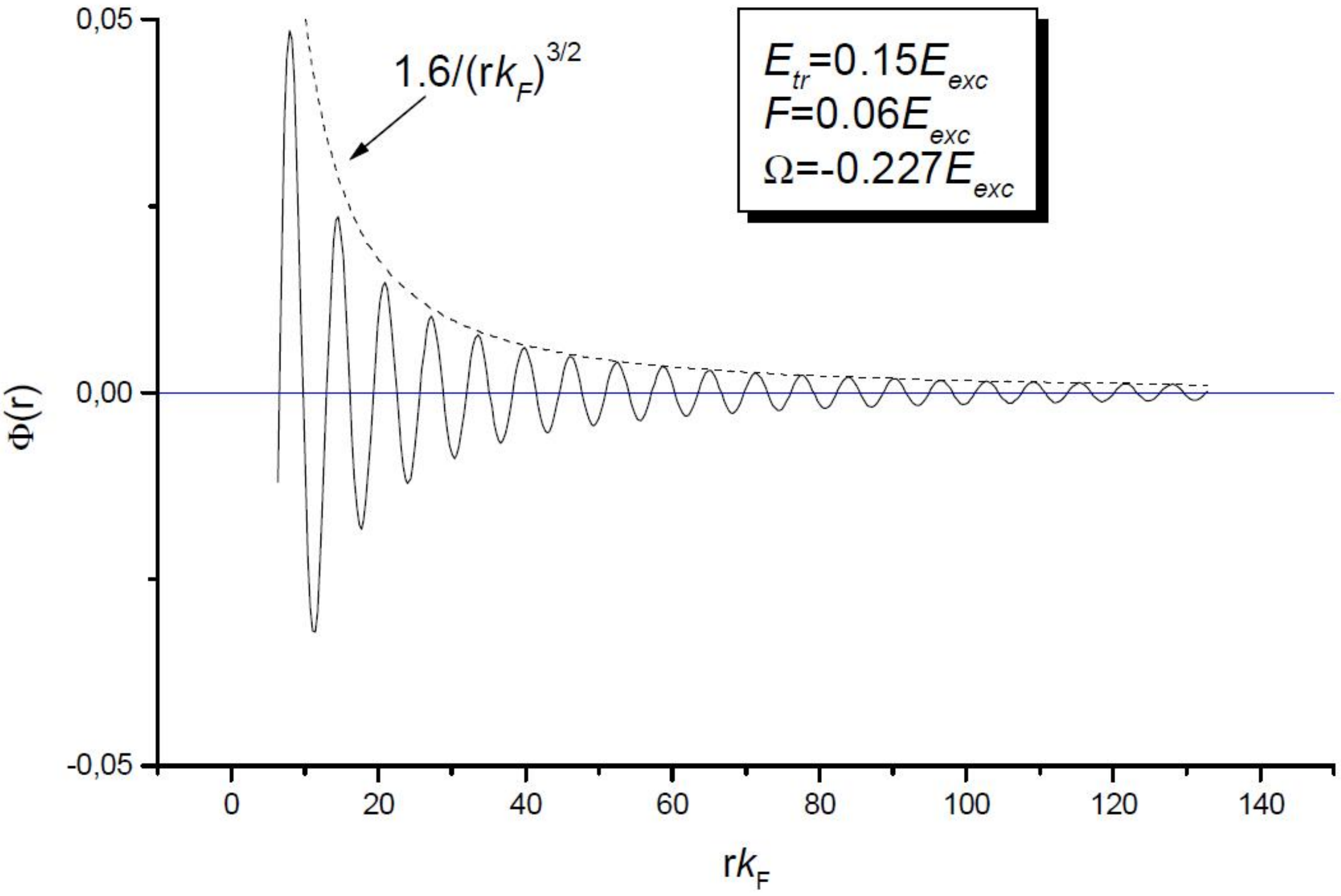

Fig. 6. *Coordinate dependence of wave function of the trion + Fermi Sea hole correlated state in the case of the Fermi Sea hole localization.*

By contrast to this behavior, the function (18) in the second region is a complex function and its real and imaginary parts decrease with distance as $|\mathbf{r}|^{-1/2}$. In order to find the function in this energy region we need to define the integral (18) adding to denominator of expression for $Tr(\mathbf{k})$ (18) an negative infinitesimal imaginary term, -*iγ*. The sign of the term should be selected to obtain the hole running out from the photo-excited trion[2]. *The function (18) for the second energy region corresponds to the delocalized hole.* These states describe the photo-excited trion and the Fermi Sea hole that is running away from the trion. The behavior of the function is shown in Fig. 7.

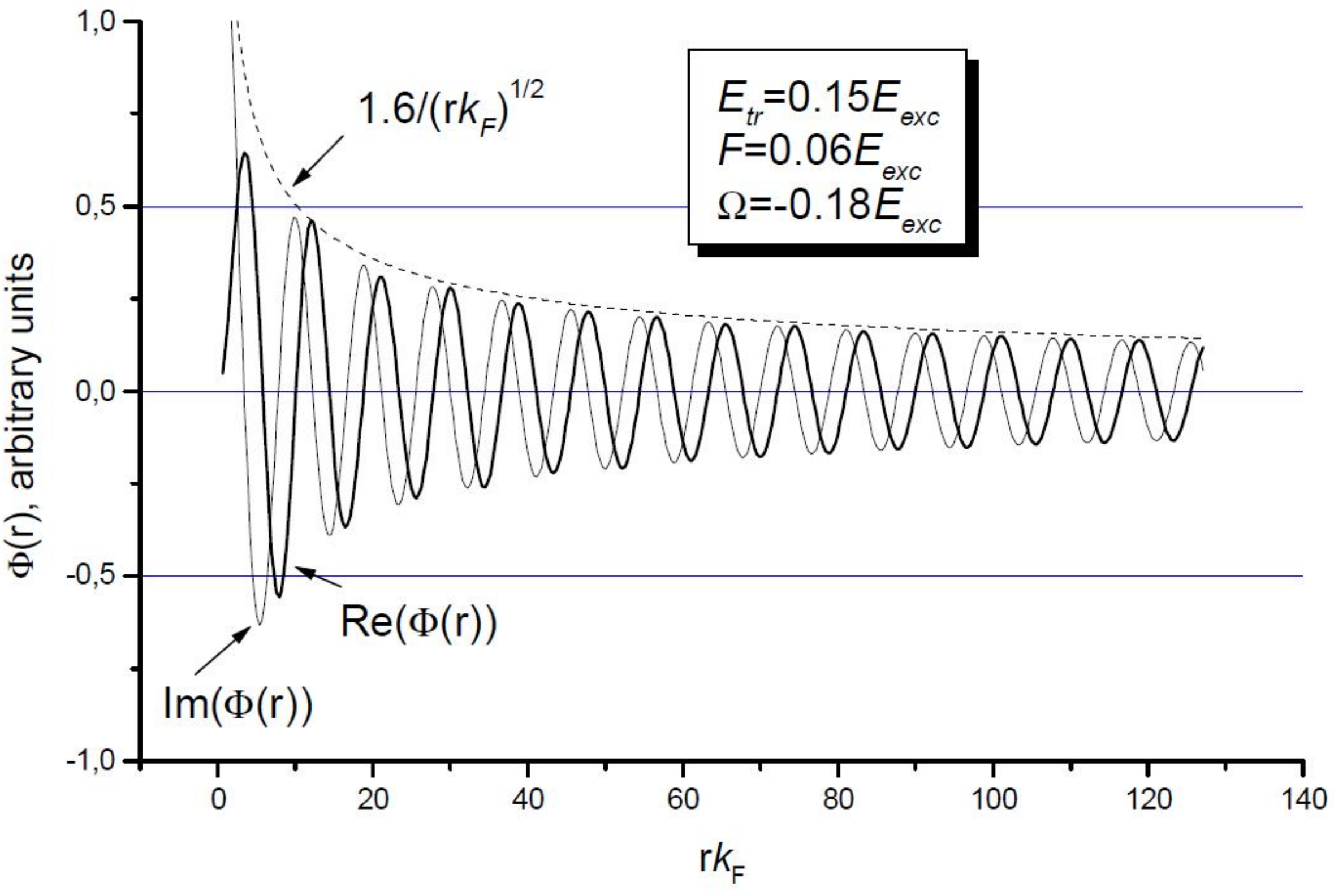

Fig.7. *Coordinate dependence of wave function of the trion + Fermi Sea hole correlated state for the delocolized Fermi Sea hole.*

[2] Generally, we can consider our system inserted in a sphere of a large radius R. In this case we have to construct our WF combining the running away and incident cylindrical waves. Normalizing of the function to unity, due to its $|\mathbf{r}|^{-1/2}$ dependence on **r** we will get the WF amplitude proportional to $1/\sqrt{R}$.

## 4. Conclusions

In vicinity of the exciton and trion resonances the process of exciton – trion mutual transformation with creation of the hole in the electron Fermi Sea plays crucial role in optics of QWs filled with free electrons.

The trion absorption spectrum consists of a band Fermi energy wide and the narrow peak at low energy side. The peak corresponds to correlated state of the trion and the hole in Fermi Sea.

The sharp trion line was observed experimentally [7]. It can be supposed to be due to the trion - Fermi Sea hole correlated state discussed above.

***Acknowledgements.*** This work was supported in part by the Russian Foundation for Basic Research (Grant No. 02-02-17610), by Program "Low-Dimensional Quantum Structures" of the Presidium of the Russian Academy of Sciences and by Programs of the Ministry of the Science, Technology and Industry of Russian Federation.

It is my pleasure to thank my colleagues, Dr. V. P. Kochereshko and Dr. D. R. Yakovlev, for very helpful and stimulating discussions.

## *Appendix*

Let us consider Schrödinger equation for Green function, *G*, of 2D electron interacting with short-range axially symmetric potential of the exciton, *V(r)*:

$$\left(\Omega-\frac{\mathbf{p}^2}{2\overline{m}}\right)G\left(\mathbf{r},\mathbf{r}'\middle|E\right)-V(r)G\left(\mathbf{r},\mathbf{r}'\middle|E\right)=\delta\left(\mathbf{r}-\mathbf{r}'\right) \qquad \text{(A.1)}$$

Scattering amplitude, $T\left(\mathbf{r},\mathbf{r}'\middle|E\right)$, is defined by the following symbolic equation:

$$G = g + gTg\,,$$

where the free-electron Green function is

$$g\left(\mathbf{r}-\mathbf{r}'\middle|\Omega\right)\equiv\int_0^\infty\frac{d^2\mathbf{k}}{(2\pi)^2}\frac{\exp\left(i\mathbf{k}\cdot(\mathbf{r}-\mathbf{r}')\right)}{\Omega-\dfrac{\mathbf{k}^2}{2\overline{m}}} \qquad \text{(A.2)}$$

Then a symbolic expansion for the scattering amplitude, *T(r,r')* is

$$T = V + VgV + VgVgV + \ldots = V + VgT \tag{A.3}$$

For δ-function-like potential $V(r) = \lambda \cdot \delta(\mathbf{r})$ with a "potential strength" $\lambda \equiv 2\pi \int_0^\infty V(r) r dr$ the solution of Eq (A.3) for $T$ is

$$(\mathbf{k}|T(E)|\mathbf{k}') = \frac{1}{\frac{1}{\lambda} - g(0|\Omega)} \tag{A.4}$$

However, the integral in Eq. (A.2) diverges on the upper limit at $|\mathbf{r} - \mathbf{r}'| \to 0$. It is due to neglect of real behavior of the potential at small distances. In order to take into account the issue, we have to replace $g(0|\Omega)$ with

$$\tilde{g}(0|\Omega) = \int_{k<\tilde{k}} \frac{d^2\mathbf{k}}{(2\pi)^2} \frac{1}{\Omega - \frac{\mathbf{k}^2}{2\overline{m}}} \tag{A.5}$$

where $\tilde{k}$ is a cut off momentum that is about the inverse radius of the potential.

It should be stressed that the scattering amplitude does not depend on $\mathbf{k}$ and $\mathbf{k}'$ as soon as the absolute values of the wave numbers are much less than the inverse potential radius.

Performing integration in Eq. 5, we have

$$\tilde{g}(0|\Omega) = \frac{\overline{m}}{2\pi} Ln\left(\frac{-\Omega}{\tilde{E}}\right).$$

Here the cut off energy is $\tilde{E} = \frac{\tilde{k}^2}{2\overline{m}}$. Obtaining the expression for $\tilde{g}(0|\Omega)$, we supposed the actual energy values, Ω, being much less than $\tilde{E}$.

Therefore we have for the scattering amplitude

$$(\mathbf{k}|T(\Omega)|\mathbf{k}') = T(\Omega) = \frac{2\pi}{\overline{m}} \frac{1}{Ln\left(\frac{\tilde{E} e^{2\pi/m\lambda}}{-\Omega}\right)} \tag{A.6}$$

For attracting potential (singlet state of trion in our case) we have $\lambda < 0$ and the scattering amplitude has a pole at small negative energy:

$$\Omega_0 \equiv -E_{binding} = -\tilde{E} e^{-2\pi/\overline{m}|\lambda|} \tag{A7}$$

The pole gives binding energy of the particle bound state in potential *V*. The result is valid for $E_{binding} << \tilde{E}$ e.g. for $\frac{|\lambda|\overline{m}}{2\pi} < 1$. In the case of trion in singlet state, the characteristic length, $\tilde{k}^{-1}$, is of order of the exciton radius and the characteristic energy, $\tilde{E}$, is of order of the exciton binding energy, $E_{exc}$ and $E_{binding} = E_{tr}$. Therefor, the first of equations (9) for the scattering amplitude is valid as soon as the trion binding energy is much less than the exciton binding energy. Comparing Eq (A7) with Eq (7), we have that

$$e^{2\pi / \overline{m}|\lambda|} \approx E_{exc} / E_{tr} \tag{A8}$$

For repulsive potential (triplet state of trion), we have λ >0. Using Eq (A8) and Eq (A6), we obtain the second expression of (9).

Generally, the value of the parameter λ for triplet state differs from the absolute value of λ for singlet state. However, having in mind the weak logarithmic dependence of the scattering amplitude on parameters and energy, we can consider the second equation of (9) for the scattering amplitude as a reasonable approximation. It should be stressed that, independently on the value of parameters, the logarithmic behavior of the both of the scattering amplitudes (9) are true for small energy values when $\Omega << \tilde{E}$.